\documentclass[12pt]{article}
\usepackage{graphics}
\begin{document}

\date{\today}


\title{Control the entanglement of two atoms in an optical cavity via white noise}

\author{Jing-Bo Xu and Shang-Bin Li\\
Zhejiang Institute of Modern Physics and Department of Physics,\\
Zhejiang University, Hangzhou 310027, People's Republic of
China\thanks{Mailing address}}
\date{}
\maketitle

\begin{abstract}
{\normalsize Two two-level atoms within a leaky optical cavity is
driven by two independent external optical white noise fields. We
investigate how entanglement between two atoms arises in such a
situation. The steady state entanglement of two atoms is also
investigated. A stochastic-resonance-like behavior of entanglement
is revealed. Finally, the
Bell violation between atoms is discussed.\\

PACS numbers: 03.65.Ud, 03.67.-a, 05.40.Ca}
\end{abstract}

\newpage

\section * {I. INTRODUCTION}
Quantum entanglement plays a crucial role in quantum information
and quantum computation [1]. Entanglement can exhibit the nature
of a nonlocal correlation between quantum systems that have no
classical interpretation. However, real quantum systems will
unavoidably be influenced by surrounding environments. The
interaction between the environment and quantum systems of
interest can lead to decoherence. It is therefore of great
importance to prevent or minimize the influence of environmental
noise in the practical realization of quantum information
processing. In order to prevent the effect of decoherence, several
approaches have been proposed such as quantum error-correcting
approach [2] or quantum error-avoiding approach
[3,4].\\

Instead of attempting to shield the system from the environmental
noise, Plenio and Huelge [5] use white noise to play a
constructive role and generate the controllable entanglement by
incoherent sources. They showed that the noise-assisted
entanglement exhibits the stochastic resonance behavior. Similar
work on this aspect has also been considered by other authors
[6-8]. In this paper, we study the quantum system in which two
two-level atoms within a leaky optical cavity is driven by two
independent external optical white noise fields. We investigate
how entanglement between two atoms arises in such a situation. It
is shown that white noise exhibits dual aspects, i.e., playing
either a destructive or a constructive role in quantum information
processing. Recently, Clark and Parkins [9] proposed a scheme to
controllably entangle the internal states of two atoms trapped in
a high-finesse optical cavity by employing quantum-reservoir
engineering. By making use of laser and cavity fields to drive two
separate Raman transitions between stable atomic ground states,
the two atoms is effectively coupled to a squeezed reservoir.
Phase-sensitive reservoir correlations leads to entanglement
between the atoms. Different from their scheme, we will focus here
on the problem of generating entanglement when only incoherent
sources are available and show that controllable entanglement can
arise in this situation. We show that, if two atoms are
simultaneous driven by two independent white noise field with the
same intensity, the entanglement between them is suppressed and
eventually completely destroyed by the noise. However, in another
case, in which only one atom is exposed in white noise field, the
steady state entanglement of the two atoms is non-monotonic
function of both the intensity of noise driving field and the
spontaneous decay rate. A double resonance behavior emerges.
Moreover, the threshold value of the spontaneous decay rate, below
which there is not any steady state entanglement,
increases with the intensity of noise field.\\

This paper is organized as follows: In section II, we outline the
experimental set up, in which two atoms are trapped in a optical
cavity and driven by the thermal field. The spectral width of the
thermal field is large compared to the linewidth of the atomic
transition so that its effect is that of a white noise source. we
model this system by a master equation and give a explicit
analytical solution of the time evolution density matrix. In
section III, based on the density matrix, we obtain the analytical
expression of the concurrence characterizing the entanglement between
two atoms. Both the entanglement during the time evolution and the steady
state entanglement are investigated. A conclusion is given in section IV.\\

\section * {II. THE MASTER EQUATION DESCRIBING TWO ATOMS TRAPPED IN A
OPTICAL CAVITY AND DRIVEN BY NOISE FIELD}

The system we consider here is two atom trapped in a optical
cavity. The atoms are driven by two independent thermal fields and
separated by a large enough distance that they feel no direct
dipole-dipole interaction. The cavity has a field decay $k$ and a
frequency $\omega$. The two eigenstates of the individual atom
($|0\rangle,|1\rangle$) constitute the qubit states. The master
equation for the total system density operator is ($\hbar=1$)
$$
\frac{d\rho}{dt}=-i[H,\rho]+{\mathcal{L}}_{cav}\rho+{\mathcal{L}}_{at}\rho,
\eqno{(1)}
$$
where
$$
H=\omega{a}^{\dagger}a+\frac{\omega_{0}}{2}\sum^{2}_{j=1}
(|1\rangle_{jj}\langle1|-|0\rangle_{jj}\langle0|)
+g\sum^{2}_{j=1}(a^{\dagger}|0\rangle_{jj}\langle1|+a|1\rangle_{jj}\langle0|),
\eqno{(2)}
$$
where $a$ and $a^{\dagger}$ are the annihilation and creation
operators of the cavity field with frequency $\omega$, and
$\omega_{0}$ is the transition frequency of the atoms and $g$ is
the atom-cavity coupling constant. The Liouvilleans
${\mathcal{L}}_{cav}\rho$ and ${\mathcal{L}}_{at}\rho$ are given
by
$$
{\mathcal{L}}_{cav}\rho=\kappa(2a\rho{a}^{\dagger}-a^{\dagger}a\rho-\rho{a}^{\dagger}a),
\eqno{(3)}
$$
and
$$
{\mathcal{L}}_{at}\rho=\sum^{2}_{j=1}(n^{(j)}_T+1)\Gamma^{(j)}
(2|0\rangle_{jj}\langle1|\rho|1\rangle_{jj}\langle0|-|1\rangle_{jj}\langle1|\rho-\rho|1\rangle_{jj}\langle1|)
$$
$$
+\sum^{2}_{j=1}n^{(j)}_T\Gamma^{(j)}
(2|1\rangle_{jj}\langle0|\rho|0\rangle_{jj}\langle1|-|0\rangle_{jj}\langle0|\rho-\rho|0\rangle_{jj}\langle0|)
\eqno{(4)}
$$
where $\Gamma^{(j)}$ describes the coupling strength of the $j$th
atom to the external fields and $n^{(j)}_T\Gamma^{(j)}$ is the
transition rate due to the thermal field. The spectral width of
the thermal field is large compared to the linewidth of the atomic
transition so that its effect is that of a white noise source.
Here, $n^{(j)}_T$ can be interpreted as an effective photon number
and that spontaneous decay of the atom out of the cavities is
included in this scenario via the $n^{(j)}_T+1$ term.\\

In the large detuning limit, i.e.,
$\Delta=\omega_0-\omega\gg{g}\sqrt{\bar{n}+1}$ with $\bar{n}$
being the mean photon number of the cavity field, there is no
energy exchange between the atomic system and the cavity. We can
obtain the effective Hamiltonian $H_e$ [10,11]
$$
H_e=\frac{g^2}{\Delta}[\sum^{2}_{j=1}(|1\rangle_{jj}\langle1|aa^{\dagger}
-|0\rangle_{jj}\langle0|a^{\dagger}a)+|1\rangle_{11}\langle0|\otimes|0\rangle_{22}\langle1|
+|0\rangle_{11}\langle1|\otimes|1\rangle_{22}\langle0|].
\eqno{(5)}
$$
The first and second terms describe the photon number dependent
Stark shifts, and the third term describes the dipole coupling
between the first and second atoms induced by the virtual photon
process. When the cavity mode is initially in the vacuum state
$|0\rangle$, it will remain in the vacuum state throughout the
procedure, the effective Hamiltonian reduces to
$$
\tilde{H}_e=\frac{g^2}{\Delta}(|1\rangle_{11}\langle1|+|1\rangle_{22}\langle1|
+|1\rangle_{11}\langle0|\otimes|0\rangle_{22}\langle1|
+|0\rangle_{11}\langle1|\otimes|1\rangle_{22}\langle0|).
\eqno{(6)}
$$
As the cavity mode will then never be populated, we can disregard
the cavity mode in the following. Now, the master equation (1) can
be reduced to
$$
\frac{d\rho_s}{dt}=-i\frac{g^2}{\Delta}(|1\rangle_{11}\langle1|+|1\rangle_{22}\langle1|
+|1\rangle_{11}\langle0|\otimes|0\rangle_{22}\langle1|
+|0\rangle_{11}\langle1|\otimes|1\rangle_{22}\langle0|)\rho_s
$$
$$
~~~~~+i\frac{g^2}{\Delta}\rho_s(|1\rangle_{11}\langle1|+|1\rangle_{22}\langle1|
+|1\rangle_{11}\langle0|\otimes|0\rangle_{22}\langle1|
+|0\rangle_{11}\langle1|\otimes|1\rangle_{22}\langle0|)
$$
$$
+\sum^{2}_{j=1}(n^{(j)}_T+1)\Gamma^{(j)}
(2|0\rangle_{jj}\langle1|\rho_s|1\rangle_{jj}\langle0|-|1\rangle_{jj}\langle1|\rho_s
-\rho_s|1\rangle_{jj}\langle1|)
$$
$$
+\sum^{2}_{j=1}n^{(j)}_T\Gamma^{(j)}
(2|1\rangle_{jj}\langle0|\rho_s|0\rangle_{jj}\langle1|-|0\rangle_{jj}\langle0|\rho_s
-\rho_s|0\rangle_{jj}\langle0|),
\eqno{(7)}
$$
where $\rho_s$ is the density matrix describing the subsystem
containing only the two atoms. Firstly, we discuss the case with
$n^{(1)}_T=n^{(2)}_T=n_T$ and $\Gamma^{(1)}=\Gamma^{(2)}=\Gamma$,
i.e., two atoms are driven by two independent thermal fields with
the same intensity. We assume that the atom 1 and atom 2 are
initially in the pure product state
$|1\rangle_{1}\otimes|0\rangle_2$. Then, the explicit analytical
solution of the master equation (7) can be obtained as follows,
$$
\rho_s(t)=\rho_{11}(t)|1\rangle_{11}\langle1|\otimes|1\rangle_{22}\langle1|
+\rho_{22}(t)|1\rangle_{11}\langle1|\otimes|0\rangle_{22}\langle0|
$$
$$
+\rho_{33}(t)|0\rangle_{11}\langle0|\otimes|1\rangle_{22}\langle1|
+\rho_{44}(t)|0\rangle_{11}\langle0|\otimes|0\rangle_{22}\langle0|
$$
$$
+\rho_{23}(t)|1\rangle_{11}\langle0|\otimes|0\rangle_{22}\langle1|
+\rho_{32}(t)|0\rangle_{11}\langle1|\otimes|1\rangle_{22}\langle0|
\eqno{(8)}
$$
with
$$
\rho_{11}(t)=\frac{n_T}{(2n_T+1)^2}[n_T+e^{-(4n_T+2)\Gamma{t}}-(n_T+1)e^{-(8n_T+4)\Gamma{t}}],
$$
$$
\rho_{22}(t)=\frac{n^2_{T}+n_T}{(2n_T+1)^2}[1+e^{-(8n_T+4)\Gamma{t}}]
+\frac{1}{2}[\frac{1}{(2n_T+1)^2}+\cos\frac{2g^2t}{\Delta}]e^{-(4n_T+2)\Gamma{t}},
$$
$$
\rho_{33}(t)=\frac{n^2_{T}+n_T}{(2n_T+1)^2}[1+e^{-(8n_T+4)\Gamma{t}}]
+\frac{1}{2}[\frac{1}{(2n_T+1)^2}-\cos\frac{2g^2t}{\Delta}]e^{-(4n_T+2)\Gamma{t}}
$$
$$
\rho_{44}(t)=\frac{n_T+1}{(2n_T+1)^2}[n_T+1-e^{-(4n_T+2)\Gamma{t}}-n_Te^{-(8n_T+4)\Gamma{t}}];
$$
$$
\rho_{23}(t)=\frac{i}{2}e^{-(4n_T+2)\Gamma{t}}\sin\frac{2g^2t}{\Delta},
$$
$$
\rho_{32}(t)=-\frac{i}{2}e^{-(4n_T+2)\Gamma{t}}\sin\frac{2g^2t}{\Delta}.
\eqno{(9)}
$$\\

In order to quantify the degree of entanglement, we adopt the
concurrence $C$ defined by Wooters [12]. The concurrence varies
from $C=0$ for an unentangled state to $C=1$ for a maximally
entangled state. For two qubits, in the "Standard" eigenbasis:
$|1\rangle\equiv|11\rangle$, $|2\rangle\equiv|10\rangle$,
$|3\rangle\equiv|01\rangle$, $|4\rangle\equiv|00\rangle$, the
concurrence may be calculated explicitly from the following:
$$
C=\max\{\lambda_1-\lambda_2-\lambda_3-\lambda_4,0\},
\eqno{(10)}
$$
where the $\lambda_{i}$($i=1,2,3,4$) are the square roots of the
eigenvalues \textit{in decreasing order of magnitude} of the
"spin-flipped" density matrix operator
$R=\rho_s(\sigma^{y}\otimes\sigma^{y})\rho^{\ast}_s(\sigma^{y}\otimes\sigma^{y})$,
where the asterisk indicates complex conjugation.\\
It is straightforward to compute analytically the concurrence for
the density matrix $\rho_s(t)$, and the concurrence $C_s(t)$
related to the density matrix $\rho_s(t)$ is obtained as follows
$$
C_s(t)=2\max\{0,|\rho_{23}(t)|-\sqrt{\rho_{11}(t)\rho_{44}(t)}\},
\eqno{(11)}
$$
where $|x|$ gives the absolute value of $x$. In Fig.1, we have
plotted the concurrence $C_s(t)$ as a function of the time $t$ and
the intensity of the thermal field $n_T$ with $g^2/\Delta=0.2$ and
$\Gamma=0.01$. It is shown that the entanglement between two atoms
decreases with $n_T$, and there is not any entanglement arising
between two atoms during the time evolution when $n_T$ is beyond a
threshold value depending on the coupling constant $\Gamma$ and
effective Rabi frequency $g^2/\Delta$. In Fig.2, the concurrence
$C_s(t)$ is plotted as the function of the time $t$ and the
coupling constant $\Gamma$ of the atoms and the external field
(Note that $\Gamma$ is equivalent to the spontaneous emission rate
if $n_T=0$) with two different values of effective photon number
$n_T$ of the thermal field. In the case with $n_T=0$ and
$g^2/\Delta=0.2$ (see Fig.2(a)), the entanglement of two atoms
always arises during the time evolution even in the presence of
atomic spontaneous emission. However, in the case with $n_T=0.3$
and $g^2/\Delta=0.2$ (see Fig.2(b)), a threshold value of $\Gamma$
is found, beyond which there is not any entanglement arising
during time evolution. All of the above discussions indicate that
the two equal intensity independent thermal fields suppress the
entanglement generation. But, that is not the full aspects
concerning the role of thermal field played in the entanglement of
two atoms. In the following section, we consider the situation in
which, only one of the atoms is driven by the external thermal
field. A different aspect of the thermal field will be found.

\section * {III. THE STEADY STATE ENTANGLEMENT OF TWO ATOMS}
In the above section, we have discussed the case, in which the two
atoms are simultaneously driven by an external thermal field.
there is not any steady state entanglement between two atoms in
that situation. In this section, we consider the situation in
which, only one of the atoms is driven by the external thermal
field. The master equation is given by
$$
\frac{d\rho_s}{dt}=-i\frac{g^2}{\Delta}(|1\rangle_{11}\langle1|+|1\rangle_{22}\langle1|
+|1\rangle_{11}\langle0|\otimes|0\rangle_{22}\langle1|
+|0\rangle_{11}\langle1|\otimes|1\rangle_{22}\langle0|)\rho_s
$$
$$
~~~~~+i\frac{g^2}{\Delta}\rho_s(|1\rangle_{11}\langle1|+|1\rangle_{22}\langle1|
+|1\rangle_{11}\langle0|\otimes|0\rangle_{22}\langle1|
+|0\rangle_{11}\langle1|\otimes|1\rangle_{22}\langle0|)
$$
$$
+(n_T+1)\Gamma
(2|0\rangle_{11}\langle1|\rho_s|1\rangle_{11}\langle0|-|1\rangle_{11}\langle1|\rho_s
-\rho_s|1\rangle_{11}\langle1|)
$$
$$
+n_T\Gamma
(2|1\rangle_{11}\langle0|\rho_s|0\rangle_{11}\langle1|-|0\rangle_{11}\langle0|\rho_s
-\rho_s|0\rangle_{11}\langle0|)
$$
$$
+\eta
(2|0\rangle_{22}\langle1|\rho_s|1\rangle_{22}\langle0|-|1\rangle_{22}\langle1|\rho_s
-\rho_s|1\rangle_{22}\langle1|),
\eqno{(12)}
$$
where $\eta$ is the spontaneous emission rate of the atom 2. We
assume that the two atoms are initially in the ground state
$|0\rangle_1\otimes|0\rangle_2$. The explicit analytical solution
of the steady state of the master equation (12) can be obtained as
follows,
$$
\rho_{st}=\rho^{s}_{11}|1\rangle_{11}\langle1|\otimes|1\rangle_{22}\langle1|
+\rho^{s}_{22}|1\rangle_{11}\langle1|\otimes|0\rangle_{22}\langle0|
$$
$$
+\rho^{s}_{33}|0\rangle_{11}\langle0|\otimes|1\rangle_{22}\langle1|
+\rho^{s}_{44}|0\rangle_{11}\langle0|\otimes|0\rangle_{22}\langle0|
$$
$$
+\rho^{s}_{23}|1\rangle_{11}\langle0|\otimes|0\rangle_{22}\langle1|
+\rho^{s}_{32}|0\rangle_{11}\langle1|\otimes|1\rangle_{22}\langle0|
\eqno{(13)}
$$
with
$$
\rho^{s}_{11}=\frac{\Omega^2\Gamma^2n^2_{T}}{
     (\Gamma+\eta+2n_{T}\Gamma)^2(\Omega^2+\Gamma\eta+2n_{T}\Gamma\eta)},
$$
$$
\rho^{s}_{22}=\frac{n_{T}\Gamma[\eta(\Gamma+\eta+2n_{T}\Gamma)^2+
\Omega^2(\Gamma+\eta+n_{T}\Gamma)]}{(\Gamma+\eta+2n_{T}\Gamma)^2(\Omega^2+\Gamma\eta+2n_{T}\Gamma\eta)},
$$
$$
\rho^{s}_{33}=\frac{\Omega^2n_{T}\Gamma(\Gamma+\eta+n_{T}\Gamma)}
{(\Gamma+\eta+2n_{T}\Gamma)^2(\Omega^2+\Gamma\eta+2n_{T}\Gamma\eta)}
$$
$$
\rho^{s}_{44}=\frac{\Gamma\eta(1+n_{T})(\Gamma+\eta+2n_{T}\Gamma)^2
+\Omega^2(\Gamma+\eta+n_{T}\Gamma)^2}
{(\Gamma+\eta+2n_{T}\Gamma)^2(\Omega^2+\Gamma\eta+2n_{T}\Gamma\eta)};
$$
$$
\rho^{s}_{23}=\frac{in_{T}\Omega\Gamma\eta}{
     (\Gamma+\eta+2n_{T}\Gamma)(\Omega^2+\Gamma\eta+2n_{T}\Gamma\eta)},
$$
$$
\rho^{s}_{32}=\frac{-in_{T}\Omega\Gamma\eta}{
     (\Gamma+\eta+2n_{T}\Gamma)(\Omega^2+\Gamma\eta+2n_{T}\Gamma\eta)},
\eqno{(14)}
$$
where $\Omega=g^2/\Delta$. The concurrence $C_{st}$ related to the
steady state $\rho_{st}$ is obtained as follows
$$
C_{st}=2\max\{0,|\rho^{s}_{23}|-\sqrt{\rho^{s}_{11}\rho^{s}_{44}}\}.
\eqno{(15)}
$$
In Fig.3, we have plotted the concurrence $C_{st}$ as a function
of the spontaneous emission rate $\eta$ and the intensity of the
thermal field $n_T$ with $g^2/\Delta=0.2$ and $\Gamma=0.1$. It is
shown that the steady state entanglement exhibit a double
stochastic-resonance-like behavior, which is similar with the
results in Ref.[5]. The double stochastic-resonance-like behavior
also emerges in Fig.4, in which $C_{st}$ is depicted as a function
of the spontaneous emission rate $\eta$ and the coupling constant
$\Gamma$ with $g^2/\Delta=0.2$ and $n_{T}=2$. In Fig.5, we plot
the $C_{st}$ as a function of $\Omega$ and the intensity of the
thermal field $n_T$ with $\Gamma=0.1$ and $\eta=0.5$. It is shown
that the threshold value of $n_T$ is strongly dependent on the
value of $\Omega$.\\

From Eq.(15), we can find the threshold values of the parameters
$\Omega$, $n_{T}$, $\Gamma$ and $\eta$, beyond which there is not
entanglement in the steady state. Some simply inequalities can be
derived as follows
$$
0<\Omega<\Omega_c=\frac{(\Gamma+\eta+2n_{T}\Gamma)\sqrt{\eta^2-\Gamma\eta-n_{T}\Gamma\eta}}
{\Gamma+\eta+n_T\Gamma},
\eqno{(16)}
$$
$$
0<n_T<n_{Tc}=\frac{\eta}{\Gamma}-1.
\eqno{(17)}
$$
The gray area in Fig.6 depicts the region where the steady state
of the two atoms is entangled in the case with $\Gamma=0.1$ and
$\eta=0.5$. In Fig.7, the concurrence $C_{st}$ is plotted as a
function of the intensity of the thermal field $n_T$ with
$\Gamma=0.01$ and $\eta=0.5$ for four different values of
$g^2/\Delta$. In Fig.8(b), we show how two atoms initially in
various different product states would eventually evolve into the
entangled steady state in the presence of the external noise
driving one of the atoms. Otherwise, in the absence of the intense
enough external noise, two atoms firstly become entangled due to
the dipole coupling induced by the virtual photon process, then,
they rapidly lose the entanglement, as shown in Fig.8(a).
\\

Finally, we attempt to discuss the nonlocality of two atoms in the
steady state. The nonlocal property of two atoms can be
characterized by the maximal violation of Bell inequality. The
most commonly discussed Bell inequality is the CHSH inequality
[13,14]. The CHSH operator reads
$$
\hat{B}=\vec{a}\cdot\vec{\sigma}\otimes(\vec{b}+\vec{b^{\prime}})\cdot\vec{\sigma}
+\vec{a^{\prime}}\cdot\vec{\sigma}\otimes(\vec{b}-\vec{b^{\prime}})\cdot\vec{\sigma},
\eqno{(18)}
$$
where $\vec{a},\vec{a^{\prime}},\vec{b},\vec{b^{\prime}}$ are unit
vectors. In the above notation, the Bell inequality reads
$$
|\langle\hat{B}\rangle|\leq2. \eqno{(19)}
$$
The maximal amount of Bell violation of a state $\rho$ is given by
[15]
$$
{\mathcal{B}}=2\sqrt{\lambda+\tilde{\lambda}}, \eqno{(20)}
$$
where $\lambda$ and $\tilde{\lambda}$ are the two largest
eigenvalues of $T_{\rho}T^{\dagger}_{\rho}$. The matrix $T_{\rho}$
is determined completely by the correlation functions being a
$3\times3$ matrix whose elements are
$(T_{\rho})_{nm}={\mathrm{Tr}}(\rho\sigma_{n}\otimes\sigma_{m})$.
Here, $\sigma_1\equiv\sigma_x$, $\sigma_2\equiv\sigma_y$, and
$\sigma_3\equiv\sigma_z$ denote the usual Pauli matrices. We call
the quantity $\mathcal{B}$ the maximal violation measure, which
indicates the Bell violation when ${\mathcal{B}}>2$ and the
maximal violation when ${\mathcal{B}}=2\sqrt{2}$. For the density
operator $\rho_{st}$ in Eq.(13) characterizing the steady state of
two atoms, $\lambda+\tilde{\lambda}$ can be written as follows
$$
\lambda+\tilde{\lambda}=4|\rho^{s}_{23}|^2+
\max[4|\rho^{s}_{23}|^2,(\rho^{s}_{11}+\rho^{s}_{44}-\rho^{s}_{22}-\rho^{s}_{33})^2].
\eqno{(21)}
$$
Recently, Verstraete et al. investigated the relations between the
violation of the CHSH inequality and the concurrence for systems
of two qubits [16]. They showed that the maximal value of
${\mathcal{B}}$ for given concurrence $C$ is $2\sqrt{1+C^2}$,
which can be achieved by the pure states and some Bell diagonal
states. If the given concurrence $C$ is larger than
$\frac{\sqrt{2}}{2}$, the minimal value of ${\mathcal{B}}$ is
$2\sqrt{2}C$, which can be achieved by the maximal entangled mixed
state. Furthermore, the entangled two qubits state with the
concurrence $C\leq\frac{\sqrt{2}}{2}$ may not violate any CHSH
inequality, even after all possible local filtering operations,
except their Bell diagonal normal form does violate the CHSH
inequalities [16]. So, it is not difficult to understand the
following results. Our calculations show that not any violation of
CHSH inequality will be found in the steady state even though the
steady state is entangled. Moreover, the stochastic-resonance-like
behavior can not be observed in the Bell violation of two atoms
during the evolution, and the stronger the noise intensity, the
more rapid the Bell violation disappears, which is shown in Fig.9.

\section * {V. CONCLUSSION}
In this paper, we investigate the problem of generating
entanglement when only incoherent sources are available and show
that controllable entanglement can arise in this situation. We
show that, if two atoms are simultaneous driven by two independent
white noise field with the same intensity, the entanglement
between them is suppressed and eventually completely destroyed by
the noise. However, in another case, in which only one atom is
exposed in white noise field, the steady state entanglement of the
two atoms is non-monotonic function of both the intensity of noise
driving field and the spontaneous decay rate. A double resonance
behavior emerges. Moreover, the threshold value of the spontaneous
decay rate, below which there is not any steady state
entanglement, increases with the intensity of noise field.\\

\section*{ACKNOWLEDGMENT}
This project was supported by the National Natural Science
Foundation of China (Project NO. 10174066).

\newpage
{\Large\bf Figure Caption}
\begin{description}
\item[FIG.1]the concurrence $C_s(t)$ is depicted as a function of the time $t$ and
the intensity of the thermal field $n_T$ with $g^2/\Delta=0.2$ and
$\Gamma=0.01$.\\
\item[FIG.2]the concurrence $C_s(t)$ is plotted as the function of the time $t$ and the
coupling constant $\Gamma$ of the atoms and the external field
 with $g^2/\Delta=0.2$ and two different values of effective photon number
$n_T$ of the thermal field, (a) $n_T=0$; (b)
$n_T=0.3$. \\
\item[FIG.3]the concurrence $C_{st}$ is plotted as a function
of the spontaneous emission rate $\eta$ and the intensity of the
thermal field $n_T$ with $g^2/\Delta=0.2$ and $\Gamma=0.1$.\\
\item[FIG.4]The concurrence $C_{st}$ depicted as a function
of the spontaneous emission rate $\eta$ and the coupling constant
$\Gamma$ with $g^2/\Delta=0.2$ and $n_{T}=2$.\\
\item[FIG.5]The concurrence $C_{st}$ is plotted
as a function of $\Omega$ and the intensity of the
thermal field $n_T$ with $\Gamma=0.1$ and $\eta=0.5$.\\
\item[FIG.6]This figure depicts the region where the steady state of the two
atoms is entangled in the case with $\Gamma=0.1$ and $\eta=0.5$.\\
\item[FIG.7]The concurrence $C_{st}$ is plotted
as a function of the intensity of the thermal field $n_T$ with
$\Gamma=0.01$ and $\eta=0.5$ for four different values of
$g^2/\Delta$ (from top to bottom,
$g^2/\Delta=0.49$, $g^2/\Delta=0.50$, $g^2/\Delta=0.505$ and $g^2/\Delta=0.51$).\\
\item[FIG.8]The concurrence $C$ is plotted
as a function of the time $t$ with $\Gamma=0.1$, $g^2/\Delta=0.2$
and $\eta=0.5$ for two different values $n_T=10^{-6}$ (a) and
$n_{T}=2$ (b) of intensity of the thermal field, and for three
different initial states: (Solid line)
$|0\rangle_1\otimes|0\rangle_2$; (Dot line)
$|1\rangle_1\otimes|0\rangle_2$; (Dash Dot line)
$|0\rangle_1\otimes|1\rangle_2$ (The values of the solid line in
(a) is too small to be seen).\\
\item[FIG.9]The maximal violation ${\mathcal{B}}$ is plotted
as a function of the time $t$ with $\Gamma=0.01$, $g^2/\Delta=0.2$
and $\eta=0.01$ for three different values of $n_T$, $n_T=0$
(Solid line), $n_T=0.5$ (Dot line), $n_T=1$ (Dash Dot line).
Two atoms are initially prepared in $|1\rangle_1\otimes|0\rangle_2$.\\
\end{description}

\begin{figure}
\centering
\includegraphics{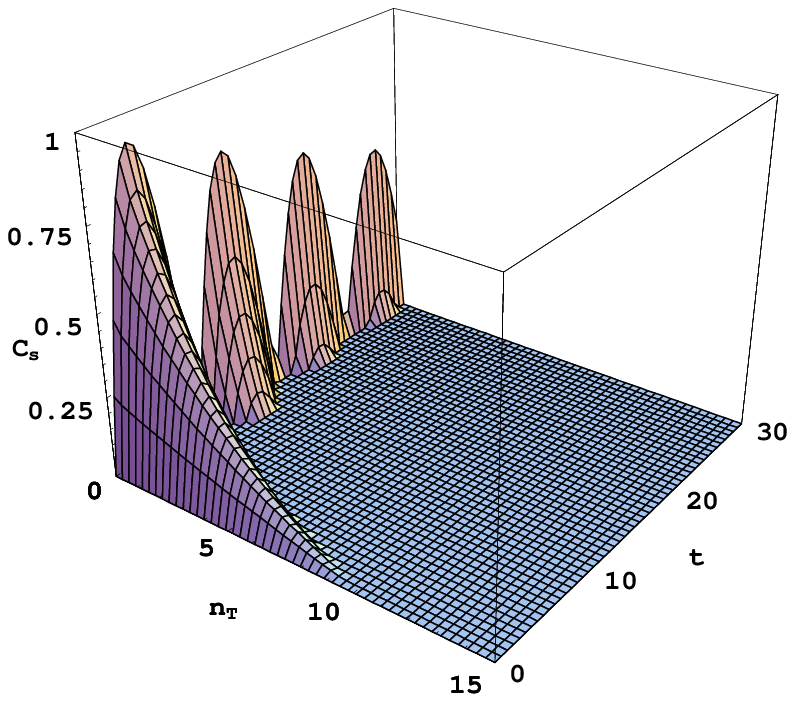}
\caption{the concurrence $C_s(t)$ is depicted as a function of the
time $t$ and the intensity of the thermal field $n_T$ with
$g^2/\Delta=0.2$ and $\Gamma=0.01$. \label{Fig.1}}
\end{figure}

\begin{figure}
\centering
\includegraphics{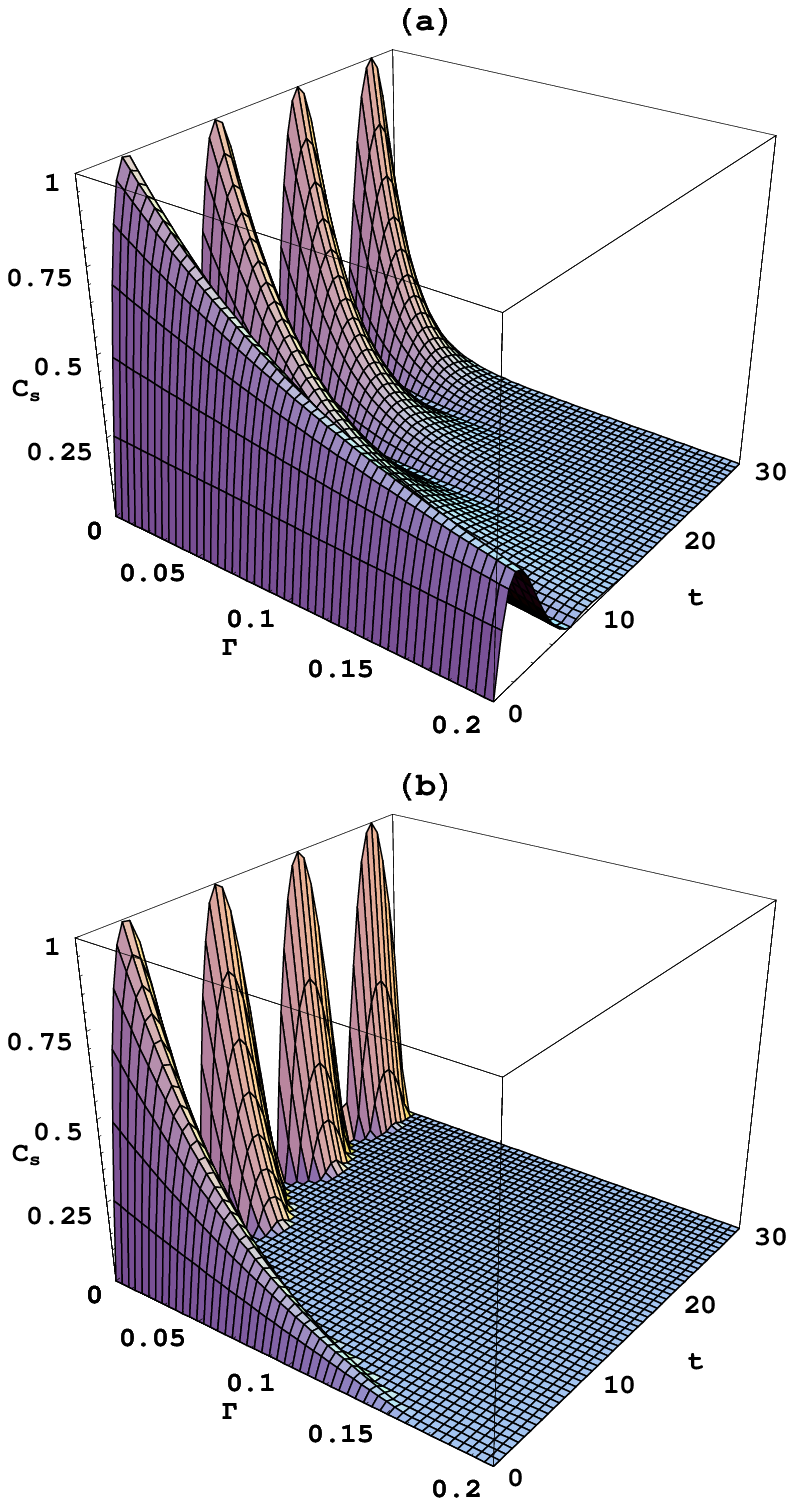}
\caption{the concurrence $C_s(t)$ is plotted as the function of
the time $t$ and the coupling constant $\Gamma$ of the atoms and
the external field
 with $g^2/\Delta=0.2$ and two different values of effective photon number
$n_T$ of the thermal field, (a) $n_T=0$; (b) $n_T=0.3$.
\label{Fig.2}}
\end{figure}

\begin{figure}
\centering
\includegraphics{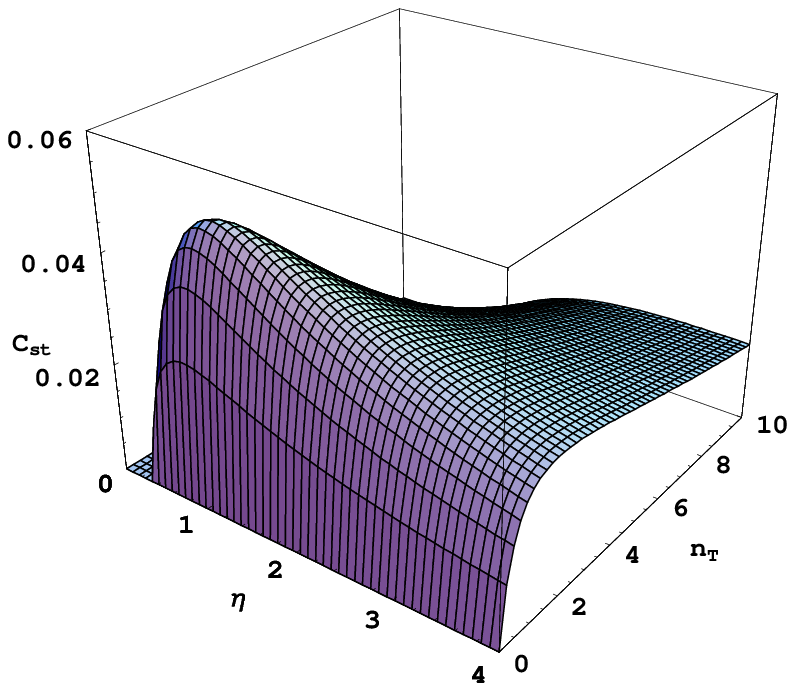}
\caption{the concurrence $C_{st}$ is plotted as a function of the
spontaneous emission rate $\eta$ and the intensity of the thermal
field $n_T$ with $g^2/\Delta=0.2$ and $\Gamma=0.1$. \label{Fig.3}}
\end{figure}

\begin{figure}
\centering
\includegraphics{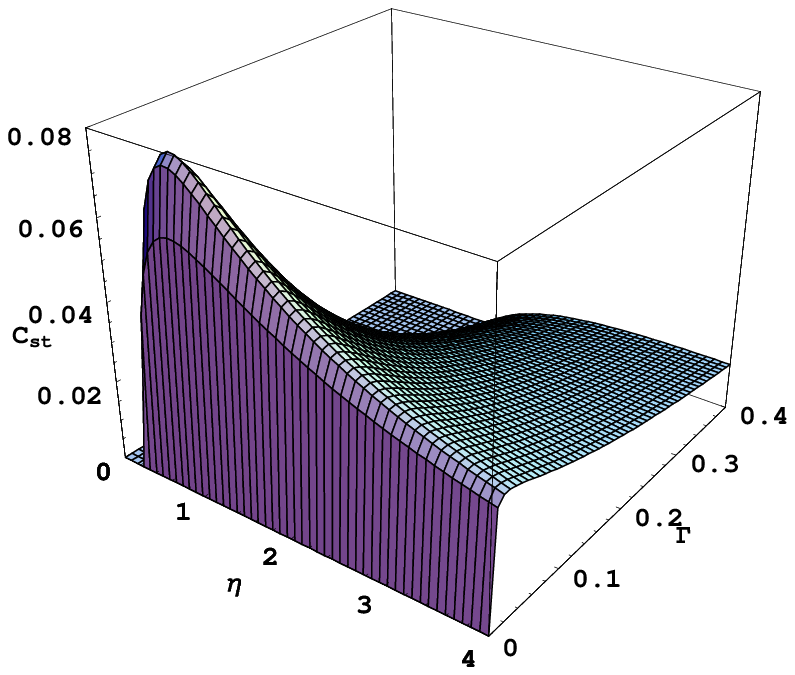}
\caption{The concurrence $C_{st}$ depicted as a function of the
spontaneous emission rate $\eta$ and the coupling constant
$\Gamma$ with $g^2/\Delta=0.2$ and $n_{T}=2$. \label{Fig.4}}
\end{figure}

\begin{figure}
\centering
\includegraphics{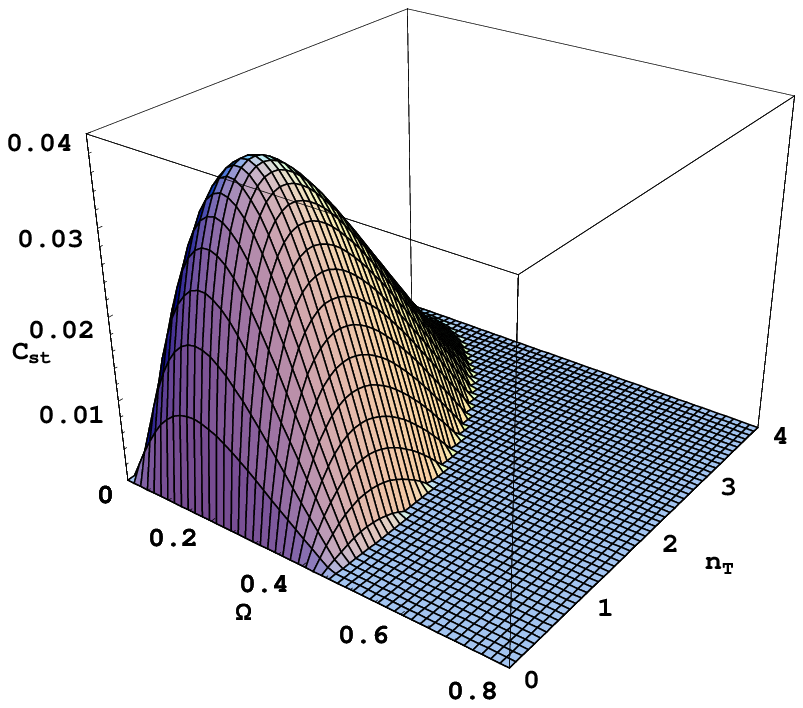}
\caption{The concurrence $C_{st}$ is plotted as a function of
$\Omega$ and the intensity of the thermal field $n_T$ with
$\Gamma=0.1$ and $\eta=0.5$. \label{Fig.5}}
\end{figure}

\begin{figure}
\centering
\includegraphics{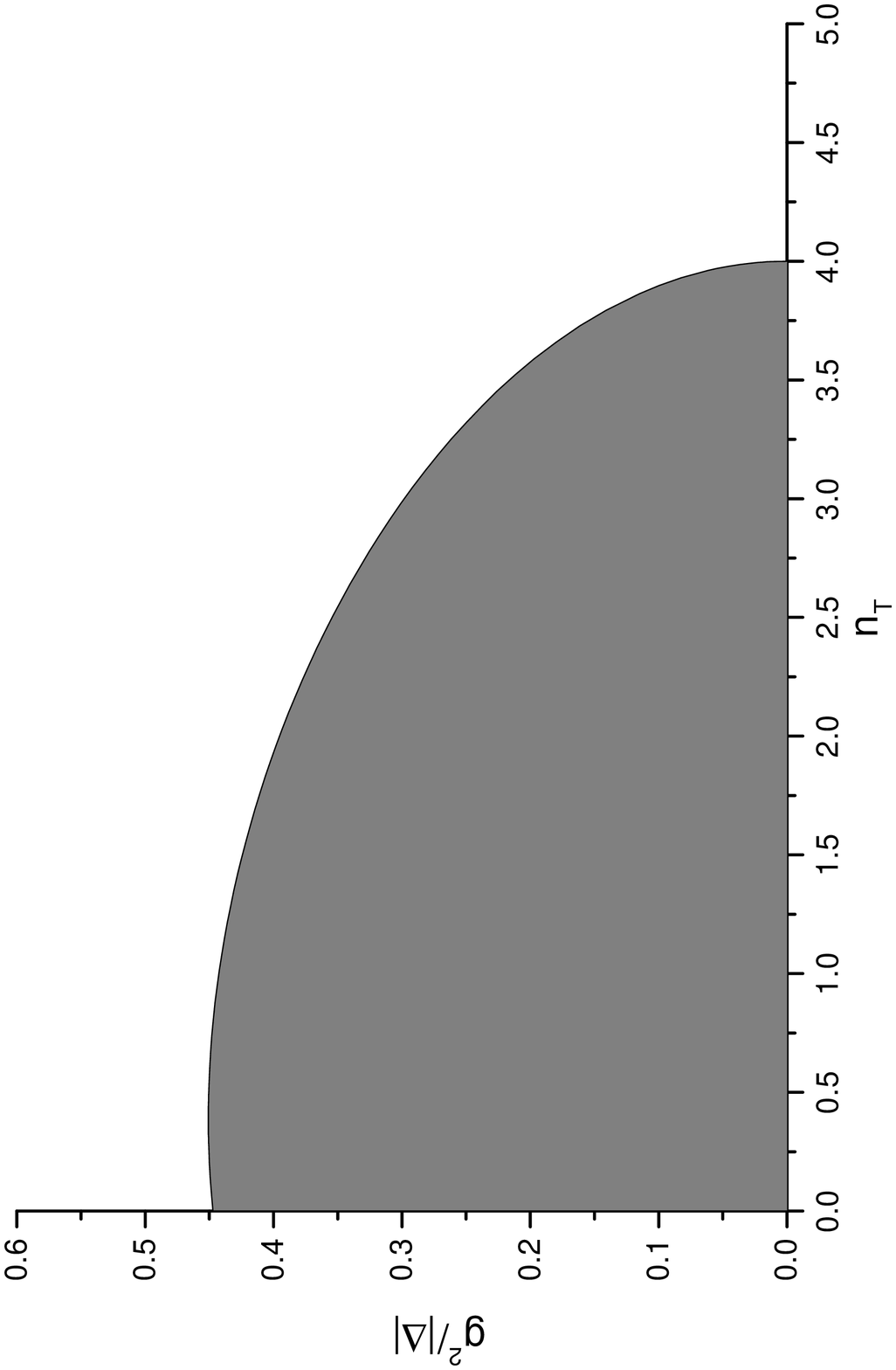}
\caption{This figure depicts the region where the steady state of
the two atoms is entangled in the case with $\Gamma=0.1$ and
$\eta=0.5$. \label{Fig.6}}
\end{figure}

\begin{figure}
\centering
\includegraphics{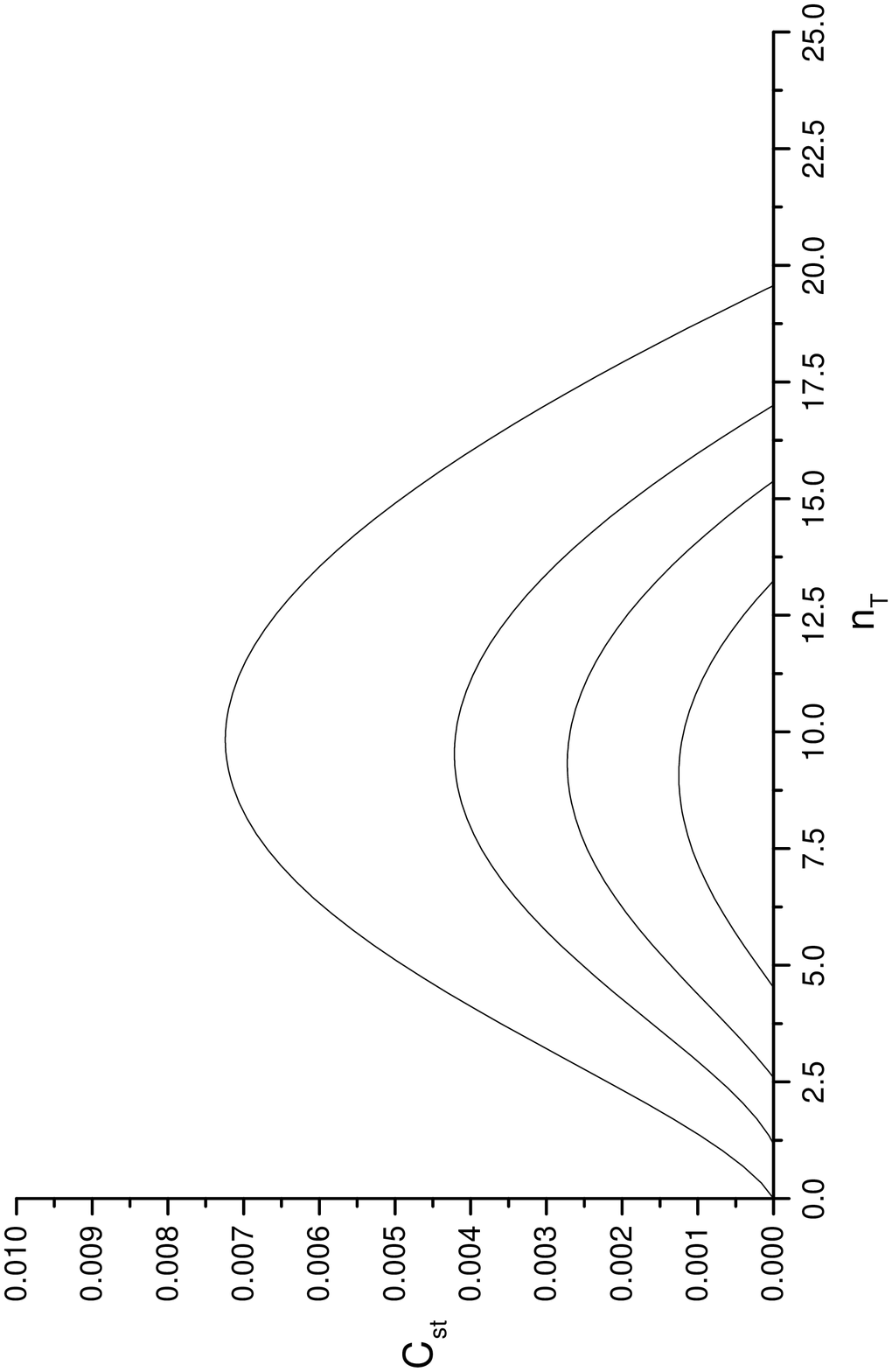}
\caption{The concurrence $C_{st}$ is plotted as a function of the
intensity of the thermal field $n_T$ with $\Gamma=0.01$ and
$\eta=0.5$ for four different values of $g^2/\Delta$ (from top to
bottom, $g^2/\Delta=0.49$, $g^2/\Delta=0.50$, $g^2/\Delta=0.505$
and $g^2/\Delta=0.51$). \label{Fig.7}}
\end{figure}

\begin{figure}
\centering
\includegraphics{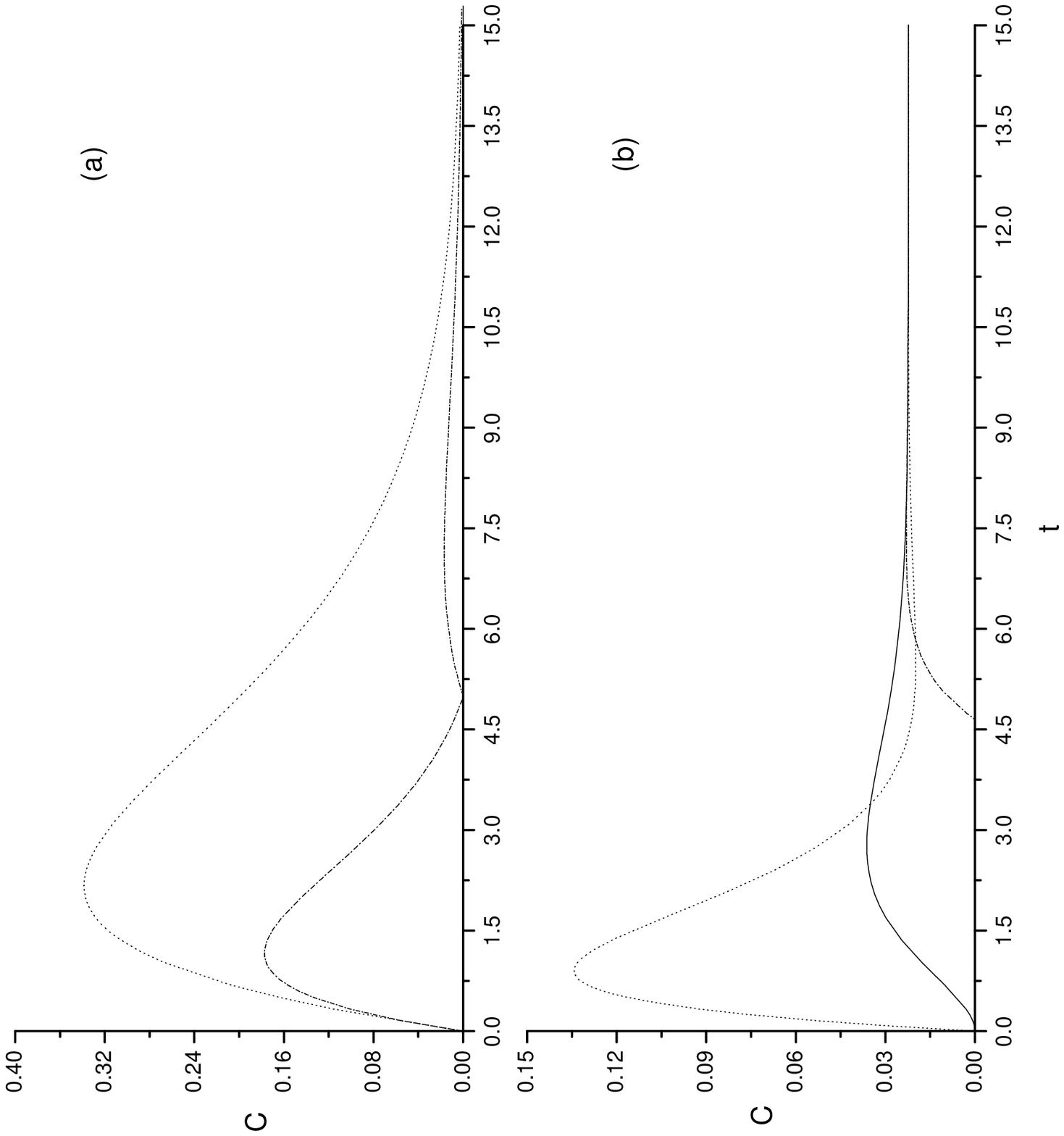}
\caption{The concurrence $C$ is plotted as a function of the time
$t$ with $\Gamma=0.1$, $g^2/\Delta=0.2$ and $\eta=0.5$ for two
different values $n_T=10^{-6}$ (a) and $n_{T}=2$ (b) of intensity
of the thermal field, and for three different initial states:
(Solid line) $|0\rangle_1\otimes|0\rangle_2$; (Dot line)
$|1\rangle_1\otimes|0\rangle_2$; (Dash Dot line)
$|0\rangle_1\otimes|1\rangle_2$ (The values of the solid line in
(a) is too small to be seen). \label{Fig.8}}
\end{figure}

\begin{figure}
\centering
\includegraphics{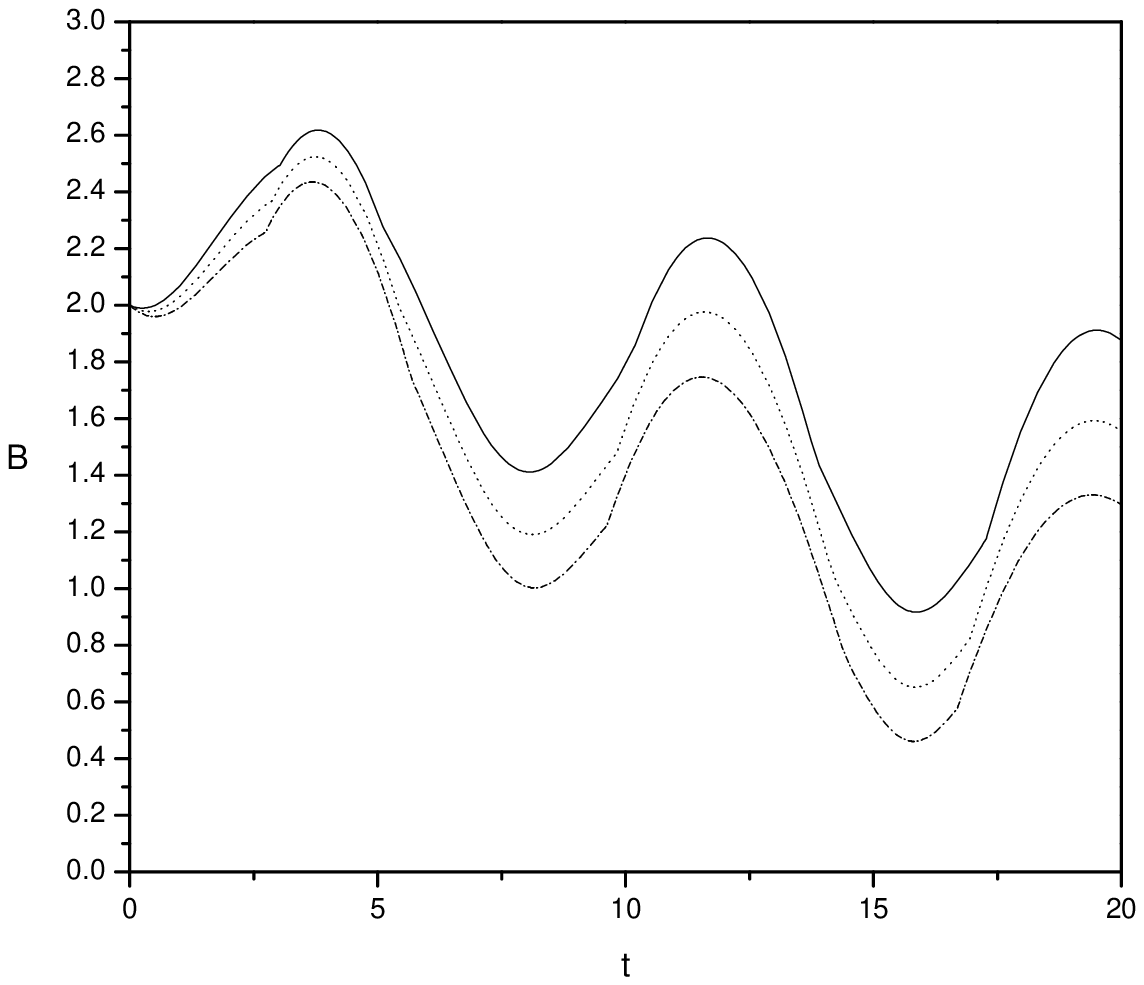}
\caption{The maximal violation ${\mathcal{B}}$ is plotted as a
function of the time $t$ with $\Gamma=0.01$, $g^2/\Delta=0.2$ and
$\eta=0.01$ for three different values of $n_T$, $n_T=0$ (Solid
line), $n_T=0.5$ (Dot line), $n_T=1$ (Dash Dot line). Two atoms
are initially prepared in $|1\rangle_1\otimes|0\rangle_2$.
\label{Fig.9}}
\end{figure}


\newpage
\begin{thebibliography}{99}
\bibitem{1}P.W. Shor, Phys. Rev. A \textbf{52}, 2493 (1995);
D.P. DiVincenzo, Science \textbf{270}, 255 (1995); L.K. Grover,
Phys. Rev. Lett. \textbf{79}(2), 325 (1997); J.I. Cirac and P.
Zoller, Nature \textit{404}, 579 (2000); M.A. Nielsen and I.L.
Chuang, \textit{Quantum Computation and Quantum Information}
(Cambridge University Press, Cambridge, 2000).
\bibitem{2}P. Zanardi, Phys. Rev. A \textbf{63}, 012301 (2001),
and references therein.
\bibitem{3}D. Kielpinski et al., Science \textbf{291}, 1013 (2001).
\bibitem{4}P.G. Kwiat et al., Science \textbf{290}, 498 (2000).
\bibitem{5}M.B. Plenio, S.F. Huelga, Phys. Rev. Lett. \textbf{88},
197901 (2002).
\bibitem{6}M. Plenio, S.F. Huelga, A. Beige, and P.L. Knight,
Phys. Rev. A \textbf{59}, 2468 (1999); S. Schneider and G.J.
Milburn, Phys. Rev. A \textbf{65}, 042107 (2002).
\bibitem{7}A. Beige, S. Bose, D. Braun, S.F. Huelga, P.L. Knight, M.B. Plenio,
 and V. Vedral, J.
Mod. Opt. \textbf{47}, 2583 (2000); M. S. Kim, J. Lee, D. Ahn,
P.L. Knight, Phys. Rev. A \textbf{65}, 040101(R) (2002).
\bibitem{8}C. Cabrillo, J.I. Cirac, P. Garc\'{i}a-Fern\'{a}ndez,
and P. Zoller, Phys. Rev. A \textbf{59}, 1025 (1999).
\bibitem{9}S. G. Clark and A. S. Parkins, Phys. Rev. Lett. \textbf{90},
047905 (2003).
\bibitem{10}S.B. Zheng and G.C. Guo, Phys. Rev. Lett. \textbf{85}, 2392 (2000).
\bibitem{11}E. Solano, G. S. Agarwal, and H.
Walther, Phys. Rev. Lett. \textbf{90}, 027903 (2003).
\bibitem{12}W. K. Wootters, Phys. Rev. Lett. \textbf{80}, 2245 (1998).
\bibitem{13}J.S. Bell, Physics (N. Y.) \textbf{1}, 195 (1965).
\bibitem{14}J.F. Clauser, M.A. Horne, A. Shimony, and R.A. Holt,
Phys. Rev. Lett. \textbf{23}, 880 (1969).
\bibitem{15}M. Horodecki, P. Horodecki, and R. Horodecki, Phys.
Lett. A \textbf{200}, 340 (1995).
\bibitem{16}F. Verstrete and M.M. Wolf, Phys. Rev. Lett. \textbf{89}, 170401 (2002).

\end{thebibliography}
\end{document}